\documentclass[fleqn,10pt,amsmath,amssymb,twocolumn]{wlscirep}
\usepackage{epsfig}
\usepackage{color}
\usepackage{graphicx}
\usepackage{dcolumn}
\usepackage{bm}
\usepackage{tikz}

\newcommand{\ket}[1]{|#1\rangle}

\providecommand{\openone}{\leavevmode\hbox{\small1\kern-3.8pt\normalsize1}}

\title{Cavity-based architecture to preserve quantum coherence and entanglement}

\author[1]{Zhong-Xiao Man}
\author[1]{Yun-Jie Xia}
\author[2,3,4,*]{Rosario Lo Franco}
\affil[1]{Shandong Provincial Key Laboratory of Laser Polarization
and Information Technology, Department of Physics, Qufu Normal University, Qufu 273165, China}
\affil[2]{Dipartimento di Fisica e Chimica, Universit\`{a} di Palermo, via Archirafi 36, 90123 Palermo, Italy}
\affil[3]{Instituto de F{\'{i}}sica de S{\~{a}}o Carlos, Universidade de S{\~{a}}o Paulo, CP 369, 13560-970 S{\~{a}}o Carlos, SP, Brasil}
\affil[4]{School of Mathematical Sciences, The University of Nottingham, University Park, Nottingham NG7 2RD, United Kingdom}
\affil[*]{rosario.lofranco@unipa.it}

\begin{abstract}
Quantum technology relies on the utilization of resources, like quantum coherence and entanglement, which allow quantum information and computation processing. This achievement is however jeopardized by the detrimental effects of the environment surrounding any quantum system, so that finding strategies to protect quantum resources is essential. Non-Markovian and structured environments are useful tools to this aim. Here we show how a simple environmental architecture made of two coupled lossy cavities enables a switch between Markovian and non-Markovian regimes for the dynamics of a qubit embedded in one of the cavity. Furthermore, qubit coherence can be indefinitely preserved if the cavity without qubit is perfect. We then focus on entanglement control of two independent qubits locally subject to such an engineered environment and discuss its feasibility in the framework of circuit quantum electrodynamics. With up-to-date experimental parameters, we show that our architecture allows entanglement lifetimes orders of magnitude longer than the spontaneous lifetime without local cavity couplings. This cavity-based architecture is straightforwardly extendable to many qubits for scalability.
\end{abstract}
\begin{document}

\date{\today}

\flushbottom
\maketitle
%
%
\thispagestyle{empty}

\section*{Introduction}
Entangled states are not only an existing natural form of compound systems in the quantum world, but also a basic
resource for quantum information technology \cite{nielsenchuang,benenti,amico2008RMP}. Due to the unavoidable coupling of a quantum system to the surrounding environment, quantum entanglement is subject to decay and can even vanish abruptly, a phenomenon known as early-stage disentanglement or entanglement sudden death \cite{yueberlyPRL2004,yueberlyPRL2006,doddPRA,santosPRA,yu2009Science,almeida2007Science,kimble2007PRL,eberlyScience2007,sallesPRA,aolitareview}. Harnessing entanglement dynamics and preventing entanglement from disappearing until the time a quantum task can be completed is thus a key challenge towards the feasibility of reliable quantum processing \cite{obrienreview,norireview}.

So far, a lot of researches have been devoted to entanglement manipulation and protection. A pure maximally entangled state can be obtained from decohered (partially entangled mixed) states \cite{bennett1,bennett2,panNature,kwiatNature,dongNatPhys} provided that there exist a large number of identically decohered states, which however will not work if the entanglement amount in these states is small. In situations where several particles are coupled to a common environment and the governing Hamiltonian is highly symmetric, there may appear a decoherence-free subspace that does not evolve in time \cite{zanardiPRL1997,lidarPRL,kwiatScience}: however, in this decoherence-free subspace only a certain kind of entangled state can be decoupled from the influence of the environment \cite{Zeno1,Zeno2}. 
The quantum Zeno effect \cite{Zeno3} can also be employed to manipulate decoherence process but, to prevent considerable degradation
of entanglement, special measurements should be performed very frequently at equal time intervals \cite{Zeno1,Zeno2}. 
By encoding each physical qubit of a manyqubit system onto a logical one comprising several physical qubits \cite{shorPRA,steanePRL,steaneRoySoc,calderbank,sainzPRA}, an appropriate reversal procedure can be applied to correct the error induced by decoherence after a multiqubit measurement that learns what error possibly occurred. Yet, as has been shown \cite{sainzPRA}, in some cases this method can indeed delay entanglement degradation but in other cases it leads to sudden disentanglement for states that otherwise disentangle only asymptotically. The possibility to preserve entanglement via dynamical decoupling pulse sequences has been also theoretically investigated recently for finite-dimensional or harmonic quantum environments \cite{muhktar2010PRA1,muhktar2010PRA2,wang2011PRA,pan2011JPB} and for solid state quantum systems suffering random telegraph or $1/f$ noise \cite{lofrancoPRB,lofrancospinecho}, but these procedures can be demanding from a practical point of view.

In general, environments with memory (so-called non-Markovian) suitably structured constitute a useful tool for protecting quantum superpositions and therefore the entanglement of composite systems \cite{yu2009Science,lofrancoreview,non-Mar2,non-Mar3}. It is nowadays well-known that independent qubits locally interacting with their non-Markovian environments can exhibit revivals of entanglement, both spontaneously during the dynamics \cite{lofrancoreview,bellomo2007PRL,bellomo2008PRA,lofranco2012PRA,LoFrancoNatCom} and on-demand by local operations \cite{darrigo2012AOP,orieux2015}. These revivals, albeit prolonging the utilization time of entanglement, however eventually decay. 
In several situations, the energy dissipations of individual subsystems of a composite system are responsible for disentanglement. Therefore, methods that can trap system excited-state population would be effective for entanglement preservation. 
A stationary entanglement of two independent atoms can be in principle achieved in photonic crystals or photonic-band-gap materials \cite{bellomo2008trapping,bellomo2010PhysScrManiscalco} if they are structured so as to inhibit spontaneous emission of individual atoms.
This spontaneous emission suppression induced by a photonic crystal has been so far verified experimentally for a single quantum dot \cite{expPBG} and its practical utilization for a multi-qubit assembly appears far from being reached.
Quantum interference can also be exploited to quench spontaneous emission in atomic systems \cite{interf1,interf2} and hence used to protect two-atom entanglement provided that three levels of the atoms can be used \cite{interf3}. 
Since the energy dissipations originate from excited state component of an entangled state,
a reduction of the weight of excited-state by prior weak measurement on the system before interacting with the environment followed by a reversal measurement after the time-evolution proves to be an efficient strategy to enhance the entanglement \cite{Kim,Man1,Man2}. However, the success of this measurement-based strategy is always conditional (probability less than one) \cite{Kim,Man1,Man2}.
It was shown that steady-state entanglement can be generated if two qubits share a common environment \cite{pianiPRL,Zeno1}, interact each other \cite{matteoEPJD} and are far from thermal equilibrium \cite{brunnerarxiv,plenio2002PRL,hartmannNJP,brunoNJP,brunoEPL}.
It has been also demonstrated that non-Markovianity may support the formation of stationary entanglement in a non-dissipative pure dephasing environment provided that the subsystems are mutually coupled \cite{non-Mar1}. 

Separated, independent two-level quantum systems at thermal equilibrium, locally interacting with their own environments, are however the preferable elements of a quantum hardware in order to accomplish the individual control required for quantum information processing \cite{obrienreview,norireview}. Therefore, proposals of strategies to strongly shield quantum resources from decay are essential within such a configuration. Here we address this issue by looking for an environmental architecture as simple as possible which is able to achieve this aim and at the same time realizable by current experimental technologies.  
In particular, we consider a qubit embedded in a cavity which is in turn coupled to a second cavity and show that this basic structure is able to enable transitions from Markovian to non-Markovian regimes for the dynamics of the qubit just by adjusting the coupling between the two cavities. 
Remarkably, under suitable initial conditions, this engineered environment is able to efficiently preserve qubit coherence and, when extended to the case of two noninteracting separated qubits, quantum entanglement. We finally discuss the effectiveness of our cavity-based architecture by considering experimental parameters typical of circuit quantum electrodynamics \cite{norireview,blaisPRA}, where this scheme can find its natural implementation.

\section*{Results}
Our analysis is divided into two parts. The first one is dedicated to the single-qubit architecture which shall permit us to investigate the dynamics of quantum coherence and its sensitivity to decay. The second part treats the two-qubit architecture for exploring to which extent the time of existence of quantum entanglement can be prolonged with respect to its natural disappearance time without the proposed engineered environment.

\subsection*{Single-qubit coherence preservation}

\begin{figure}[tbp]
\centering
\includegraphics[width=0.46\textwidth]{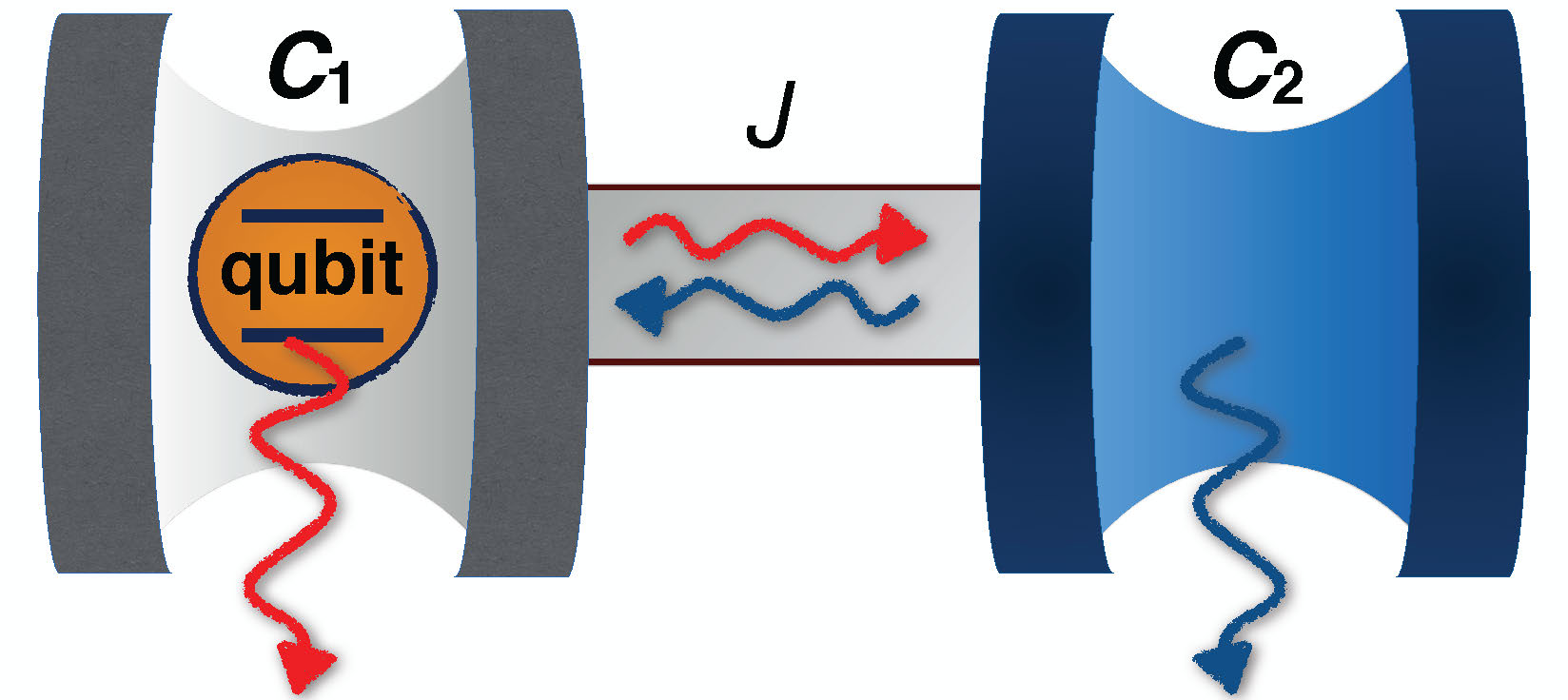}
\caption{\textbf{Scheme of the single-qubit architecture.} A two-level atom (qubit) is embedded in a cavity $C_1$ which is in turn coupled to a second cavity $C_2$ by a coupling strength $J$. Both cavities are taken at zero temperature and can lose photons.}
\label{fig:system}
\end{figure}

The global system is made of a two-level atom (qubit) inside a lossy cavity
$C_{1}$ which in turn interacts with another cavity $C_{2}$, as depicted in Fig.~\ref{fig:system}.
The Hamiltonian of the qubit and two cavities is given by ($\hbar = 1$)
\begin{eqnarray}\label{singlequbithamiltonian}
\hat{H}&=&(\omega _{0}/2)\hat{\sigma}_{z}+\omega_{1}\hat{a}_{1}^{\dag }\hat{a}_{1}
+\omega_{2}\hat{a}_{2}^{\dag }\hat{a}_{2}+\kappa(\hat{a}_{1}^{\dag}\hat{\sigma} _{-}+\hat{a}_{1}\hat{\sigma} _{+})\nonumber\\
&&+J(\hat{a}_{1}^{\dag}\hat{a}_{2}+\hat{a}_{1}\hat{a}_{2}^{\dag}),
\end{eqnarray}
where $\hat{\sigma}_{z}=\left|1\right\rangle\left\langle1\right|-\left|0\right\rangle\left\langle0\right|$
is a Pauli operator for the qubit with transition frequency $\omega _{0}$, $\hat{\sigma} _{\pm }$ are
the raising and lowering operators of the qubit, $\hat{a}_{1}$\ $(\hat{a}_{1}^{\dag })$ and
$\hat{a}_{2}$\ $(\hat{a}_{2}^{\dag })$ the annihilation (creation) operators of cavities $C_1$ and $C_2$ which sustain modes
with frequency $\omega _{1}$ and $\omega _{2}$, respectively. The parameter
$\kappa$ denotes the coupling of the qubit with cavity $C_1$ and $J$ the coupling between the two cavities.
We take $\omega _{1}=\omega _{2}=\omega$ and, in order to consider both resonant and non-resonant qubit-$C_1$ interactions, 
$\omega_{0}=\omega+\delta$ with $\delta$ being the qubit-cavity detuning. 
Taking the dissipations of the two cavities into account, the density operator $\rho(t)$ of the
atom plus the cavities obeys the following master equation \cite{petru}
\begin{eqnarray} \label{ro}
\dot{\rho}(t) &=&-i[\hat{H},\rho(t)]\nonumber\\
&-&\sum_{n=1}^{2}\frac{\Gamma _{n}}{2}[a_{n}^{\dag }a_{n}\rho(t)-2a_{n}\rho(t)a_{n}^{\dag }+\rho(t)a_{n}^{\dag }a_{n}],
\end{eqnarray}
where $\dot{\rho}(t)\equiv d\rho(t)/dt$ and $\Gamma _{1}$ ($\Gamma _{2}$) denotes the photon decay rate of cavity $C_1$ ($C_2$). The rate $\Gamma _{n}/2$ physically represents the bandwidth of the Lorentzian frequency spectral density of the cavity $C_n$, which is not a perfect single-mode cavity \cite{petru}. A cavity with a high quality factor will have a narrow bandwidth and therefore a small photon decay rate. Weak and strong coupling regimes for the qubit-$C_1$ interaction can be then individuated by the conditions $\kappa \leq \Gamma_1/4$ and $\kappa > \Gamma_1/4$ \cite{petru,bellomo2007PRL}.

Let us suppose the qubit is initially in the excited state $\left| 1\right\rangle$ and both cavities in the
vacuum states $\left| 00\right\rangle$, so that the overall initial state is $\rho(0)=\left| 100\right\rangle\left\langle 100\right|$,
where the first, second and third element correspond to the qubit, cavity $C_1$ and cavity $C_2$, respectively.
Since there exist at most one excitation in the total system at any time of evolution, we can make the ansatz for $\rho(t)$ in the form
\begin{equation}\label{rot}
\rho (t)=\left( 1-\lambda (t)\right) \left| \psi (t)\right\rangle\left\langle \psi (t)\right|
+\lambda (t)\left| 000\right\rangle\left\langle 000\right|,
\end{equation}
where $0\leq \lambda (t)\leq 1$\ with $\lambda (0)=0$ and
$\left| \psi(t)\right\rangle=h(t)\left| 100\right\rangle +c_{1}(t)\left|010\right\rangle
+c_{2}(t)\left| 001\right\rangle$
with $h(0)=1$\ and $c_{1}(0)=c_{2}(0)=0.$ It is convenient to
introduce the unnormalized state vector \cite{pseu1,pseu2}
\begin{eqnarray}  \label{unnor}
\left| \widetilde{\psi }(t)\right\rangle &\equiv &\sqrt{1-\lambda (t)}\left| \psi (t)\right\rangle \nonumber \\
&=&\widetilde{h}(t)\left| 100\right\rangle +\widetilde{c}
_{1}(t)\left| 010\right\rangle+\widetilde{c}_{2}(t)\left|001\right\rangle,
\end{eqnarray}
where $\widetilde{h}(t)\equiv \sqrt{1-\lambda (t)}h(t)$ represents the
probability amplitude of the qubit and $\widetilde{c}_{n}(t)\equiv \sqrt{1-\lambda (t)}c_{n}(t)$ ($n=1,2$)
that of the cavities being in their excited
states. In terms of this unnormalized state vector we then get
\begin{equation} \label{ronga}
\rho (t)=| \widetilde{\psi }(t)\rangle\langle
\widetilde{\psi }(t)| +\lambda (t)| 000\rangle\langle 000|.
\end{equation}
The time-dependent amplitudes $\widetilde{h}(t),$\ $\widetilde{c}_{1}(t),$\ $\widetilde{c}_{2}(t)$ of Eq. (\ref{unnor}) are determined by a set of differential equations as
\begin{eqnarray}
i\frac{d\widetilde{h}(t)}{dt} &=&(\omega+\delta)\widetilde{h}(t)+
\kappa\widetilde{c}_{1}(t), \nonumber \\
i\frac{d\widetilde{c}_{1}(t)}{dt} &=&\left( \omega -\frac{i}{2}\Gamma
_{1}\right) \widetilde{c}_{1}(t)+\kappa\widetilde{h}(t)+
J \widetilde{c}_{2}(t),  \nonumber \\
i\frac{d\widetilde{c}_{2}(t)}{dt} &=&\left( \omega -\frac{i}{2}\Gamma
_{2}\right) \widetilde{c}_{2}(t)+J \widetilde{c}_{1}(t).
\label{eqs}
\end{eqnarray}
The above differential equations can be solved by means of standard Laplace
transformations combined with numerical simulations to obtain the reduced
density operators of the atom as well as of each of the cavities.
In particular, in the basis $\{\left|1\right\rangle, \left|0\right\rangle\}$ the density matrix evolution of the qubit can be cast as
\begin{equation}\label{sing-at-evo}
\rho(t)=\left(
  \begin{array}{cc}
    u_{t}\rho_{11}(0) & z_{t}\rho^{j}_{10}(0) \\
    z_{t}^{*}\rho_{01}(0) & 1-u_{t}\rho_{11}(0) \\
  \end{array}
\right),
\end{equation}
where $u_{t}$ and $z_{t}$ are functions of the time $t$ (see Methods).

An intuitive quantification of quantum coherence is based to the off-diagonal elements of the desired quantum state, being these related to the basic property of quantum interference. Indeed, it has been recently shown \cite{coher} that the functional
\begin{equation}\label{coh}
\mathcal{C}(t)=\sum_{i,j (i\neq j)}|\varrho_{ij}(t)|,
\end{equation}
where $\varrho_{ij}(t)$\ $(i\neq j)$ are the off-diagonal elements of the system density matrix, satisfies the physical requirements which make it a proper coherence measure \cite{coher}. In the following, we adopt $\mathcal{C}(t)$ as quantifier of the qubit coherence
and explore how to preserve and even trap it under various conditions. To this aim, we first consider the
resonant atom-cavity interaction and then discuss the effects of detuning on the dynamics of coherence.

\begin{figure*}[!tbp]
\centering
{\includegraphics[width=0.6\textwidth]{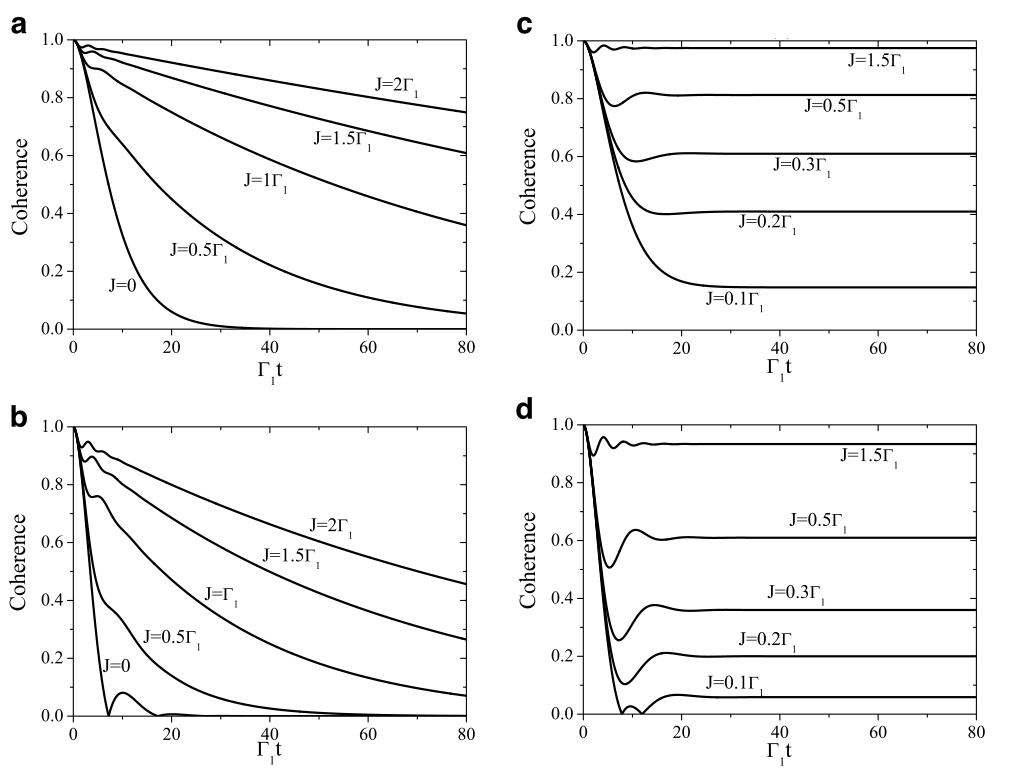}}
\caption{Coherence $\mathcal{C}(t)$ of the qubit as a function of the scaled time $\Gamma_{1}t$ for different coupling strengths $J$ between the two cavities for (a) $\kappa=0.24\Gamma_{1}$, $\Gamma_{2}=0.5\Gamma_{1}$ and (b) $\kappa=0.4\Gamma_{1}$, $\Gamma_{2}=0.5\Gamma_{1}$. The qubit is initially prepared in the state $\left|\phi(0)\right\rangle$ with $\alpha=\beta=1/\sqrt{2}$ and resonant with the cavity (detuning $\delta=0$).
The plots in panels (c) and (d) display the coherence trapping for a perfect cavity ($\Gamma_{2}=0$) with $\kappa=0.24\Gamma_{1}$ and $\kappa=0.4\Gamma_{1}$, respectively.}
\label{fig:coher}
\end{figure*}

Suppose the qubit is initially prepared in the state $\left|\phi(0)\right\rangle=
\alpha\left|0\right\rangle+\beta\left|1\right\rangle$ (with $|\alpha|^2+|\beta|^2$=1),
namely, $\mathcal{C}(0)=2|\alpha\beta|$, then at time $t>0$ the coherence becomes
$\mathcal{C}(t)=2|\alpha\beta\widetilde{h}(t)|$.
Focusing on the weak coupling between the qubit and the cavity $C_1$ with $\kappa=0.24\Gamma_{1}$,
we plot the dynamics of coherence in Fig. \ref{fig:coher}(a). In this case, the qubit exhibits a Markovian
dynamics with an asymptotical decay of the coherence in the absence of the cavity $C_2$ (with $J=0$). However, by introducing the cavity $C_2$ with a sufficiently large coupling strength, quantum coherence undergoes non-Markovian dynamics with oscillations.
Moreover, it is readily observed that the decay of coherence can be greatly inhibited by increasing the $C_1$-$C_2$ coupling strength $J$. On the other hand, if the coupling between the atom and the cavity $C_1$ is initially in the
strong regime with the occurrence of coherence collapses and revivals, the increasing of the $C_1$-$C_2$ coupling strength $J$ can drive the non-Markovian dynamics of the qubit to the Markovian one and then back to the non-Markovian one,
as shown in Fig.~\ref{fig:coher}(b). This behavior is individuated by the suppression and the successive reactivation of oscillations during the dynamics. It is worth noting that, although the qubit can experience non-Markovian dynamics again
for large enough $J$, the non-Markovian dynamics curve is different from the original one for $J=0$
in the sense that the oscillations arise before the coherence decays to zero. In general, the coupling
of $C_1$-$C_2$ can enhance the quantum coherence also in the strong coupling regime between the qubit
and the cavity $C_1$.

\begin{figure}[!tbp]
\centering
{\includegraphics[width=0.46\textwidth]{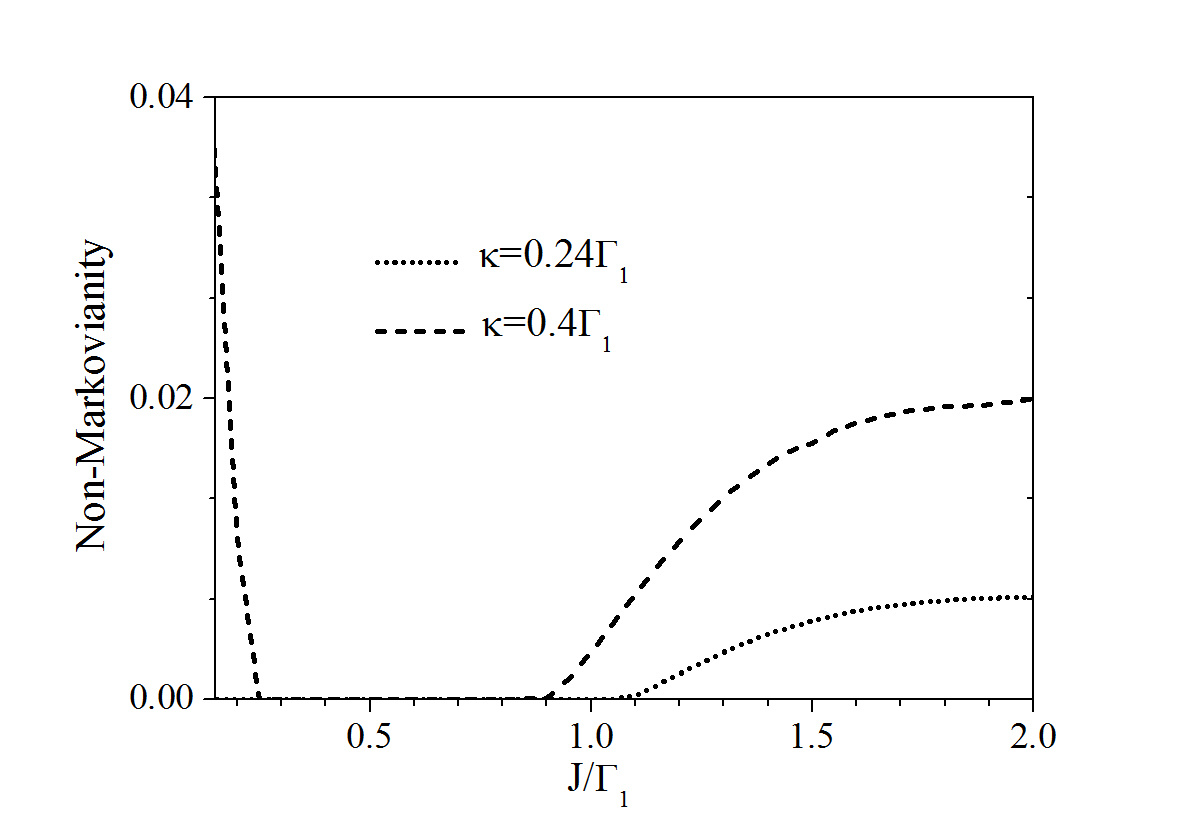}}
\caption{Non-Markovianity quantifier $\mathcal{N}$ of equation~(\ref{N}) of the qubit dynamics as a function of $J/\Gamma_1$ for weak ($\kappa = 0.24\Gamma_1$) and strong ($\kappa = 0.4\Gamma_1$) coupling regimes to cavity $C_1$ and a fixed decay rate $\Gamma_2 = 0.5\Gamma_1$ of the cavity $C_2$.}
\label{fig:nonMar}
\end{figure}

The oscillations of coherence, in clear contrast to the monotonic smooth decay in the Markovian regime, constitute a sufficient condition to signify
the presence of memory effects in the system dynamics, being due to information backflow from the environment to the quantum system \cite{breuer2009PRL}. The degree of a non-Markovian process, the so-called non-Markovianity, can be
quantified by different suitable measures
\cite{breuer2009PRL,lorenzoPRA,rivas2010PRL,bylicka2014}. We adopt here the non-Markovianity measure which exploits the dynamics of the trace distance
between two initially different states $\rho _{1}(0)$ and $\rho _{2}(0)$ of
an open system to assess their distinguishability \cite{breuer2009PRL}. A Markovian evolution can never increase the trace distance, hence nonmonotonicity of the latter would imply a non-Markovian character of the system dynamics. Based on this concept, the non-Markovianity can be
quantified by a measure $\mathcal{N}$ defined as \cite{breuer2009PRL}
\begin{equation}
\mathcal{N}=\max_{\rho _{1}(0),\rho _{2}(0)}\int_{\sigma >0}\sigma [t,\rho
_{1}(0),\rho _{2}(0)]dt,  \label{N}
\end{equation}
where $\sigma [t,\rho _{1}(0),\rho _{2}(0)]=dD[\rho _{1}(t),\rho_{2}(t)]/dt$ is the rate of change of the trace distance, which is defined as
$D[\rho _{1}(t),\rho _{2}(t)]=(1/2)\mathrm{Tr}|\rho _{1}(t)-\rho_{2}(t)|$, with $|X|=\sqrt{X^{\dag }X}.$ By virtue of $\mathcal{N}$, we plot in Fig.~\ref{fig:nonMar} the non-Markovianity of the qubit dynamics for the conditions considered in Fig. \ref{fig:coher}(a)-(b). We see that if the qubit is initially weakly coupled to the cavity $C_{1}$ ($\kappa=0.24\Gamma_{1}$) its dynamics can undergo a transition from Markovian ($\mathcal{N}=0$) to non-Markovian ($\mathcal{N}>0$) regimes by increasing the coupling strengths $J$ between the two cavities. On the other hand, for strong qubit-cavity coupling ($\kappa=0.4\Gamma_{1}$), the non-Markovian dynamics occurring for $J=0$ turns into Markovian and then back to non-Markovian by increasing $J$. We mention that such a behavior has been also observed in a different structured system where a qubit simultaneously interacts with two coupled lossy cavities \cite{lofrancoManPRA}.

\begin{figure}[!tbp]
\centering
{\includegraphics[width=0.47\textwidth]{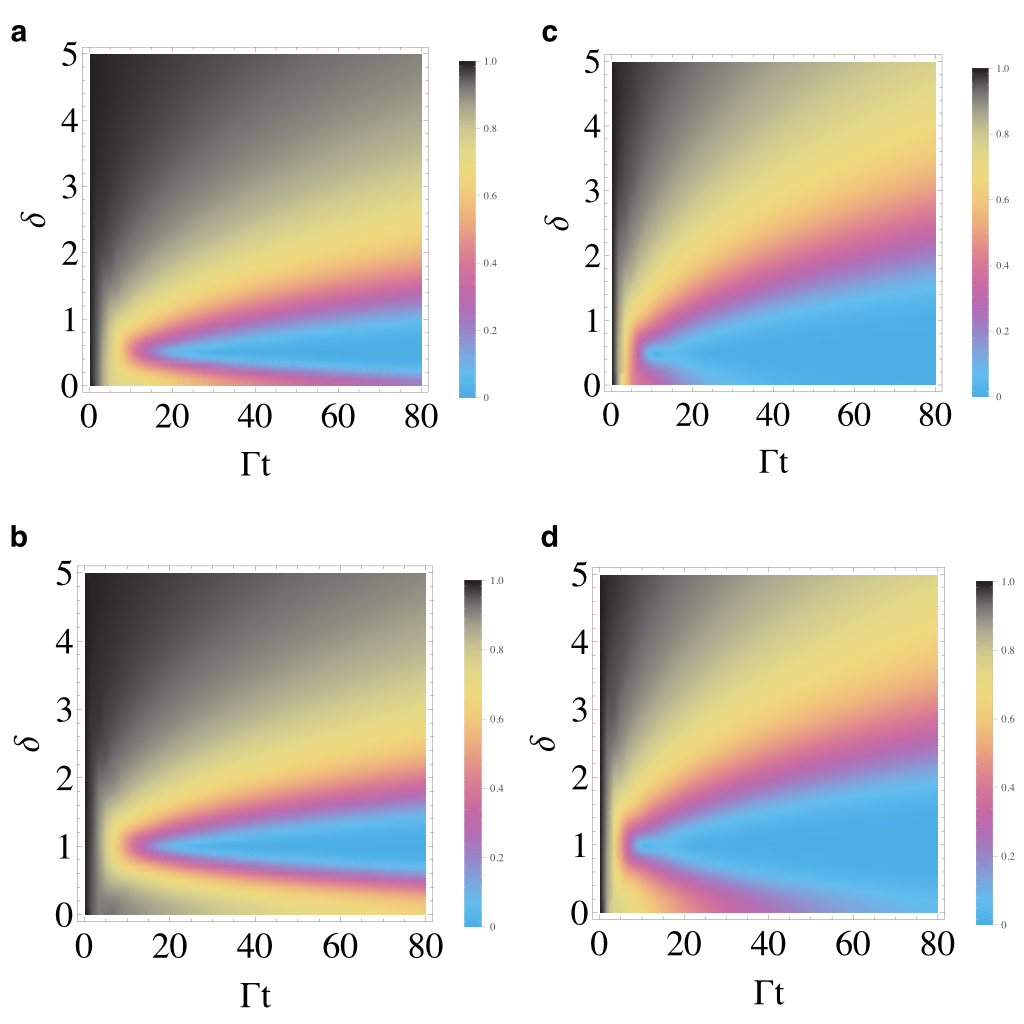}}
\caption{Density plots of coherence $\mathcal{C}(t)$ of the qubit as functions of detuning $\delta$
and the scaled time $\Gamma_{1}t$ for (a) $\kappa=0.24\Gamma_{1}$, $\Gamma_{2}=0.2\Gamma_{1}$, $J=0.5\Gamma_{1}$;
(b) $\kappa=0.24\Gamma_{1}$, $\Gamma_{2}=0.2\Gamma_{1}$, $J=\Gamma_{1}$;
(c) $\kappa=0.4\Gamma_{1}$, $\Gamma_{2}=0.5\Gamma_{1}$, $J=0.5\Gamma_{1}$;
(d) $\kappa=0.4\Gamma_{1}$, $\Gamma_{2}=0.5\Gamma_{1}$, $J=\Gamma_{1}$.
The initial state of the qubit is maximally entangled ($\alpha=\beta=1/\sqrt{2}$).
The values of the coherence are within the range: $0$ \protect\includegraphics[width=2.2 cm]{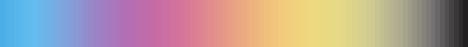} $1$.}
\label{fig:detuning}
\end{figure}

\begin{figure*}[!tbp]
\centering
{\includegraphics[width=0.8\textwidth]{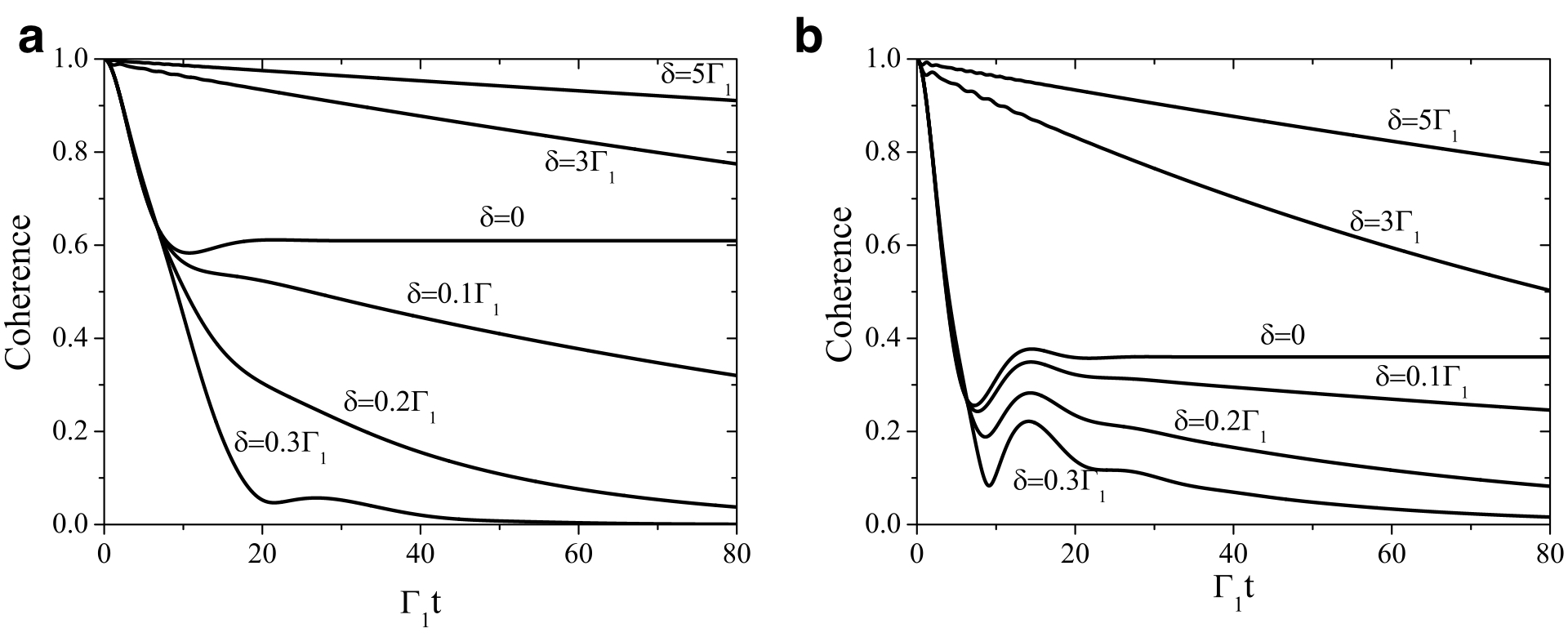}}
\caption{Coherence $\mathcal{C}(t)$ of the qubit as a function of the scaled time $\Gamma_{1}t$ for different values of the detuning $\delta$ in the case when the cavity $C_2$ is perfect, that is $\Gamma_{2}=0$. The qubit-$C_1$ and the $C_1$-$C_2$ coupling strengths are, respectively, (a) 
$\kappa=0.24\Gamma_{1}$, $J=0.3\Gamma_{1}$; (b) $\kappa=0.4\Gamma_{1}$, $J=0.3\Gamma_{1}$. Out of resonance ($\delta>0$) no coherence trapping is achievable.}
\label{fig:detuning2D}
\end{figure*}

Trapping qubit coherence in the long-time limit is a useful dynamical feature for itself that shall play a role for the preservation of quantum entanglement to be treated in the next section. We indeed find that the use of coupled cavities can achieve this result if the cavity $C_2$ is perfect, that is $\Gamma_{2}=0$ (no photon leakage). The plots in Figure \ref{fig:coher}(c)-(d) demonstrate the coherence trapping
in the long-time limit for both weak and strong coupling regimes between the qubit and the cavity $C_{1}$
for different coupling strengths $J$ between the two cavities.
This behavior can be explained by noticing that there exists a bound (decoherence-free) state of the qubit and the cavity $C_2$
of the form $\left|\psi_{-}\right\rangle=J\left|10\right\rangle-\kappa\left|01\right\rangle$,
with $J$ and $\kappa$ being the $C_1$-$C_2$ and qubit-$C_1$ coupling strengths. Being this state free from decay, once the reduced initial state
of the qubit and the cavity $C_2$ contains a nonzero component of this bound state
$\left|\psi_{-}\right\rangle$, a long-living quantum coherence for the qubit can be obtained.
For the initial state $\left|\Phi(0)\right\rangle=
\alpha\left|000\right\rangle+\beta\left|100\right\rangle$ of the qubit and two cavities here considered and $\Gamma_2=0$,
the coherence defined in Eq. (\ref{coh}) gets the asymptotic value $\mathcal{C}(t\rightarrow\infty)=2|\alpha\beta J^{2}/(J^{2}+\kappa^{2})|$,
which increases with $J$ for a given $\kappa$. We further point out that the cavity $C_1$ acts as a catalyst of the entanglement for the hybrid qubit-$C_2$ system, in perfect analogy to the stationary entanglement exhibited by two qubits embedded in a common cavity \cite{Zeno1}. In the latter case, in fact, the cavity mediates the interaction between the two qubits and performs as an entanglement catalyst for them.

We now discuss the effect of non-resonant qubit-$C_1$ interaction ($\delta\neq0$) on the dynamics of coherence. In Figure \ref{fig:detuning}(a)-(d), we display the density plots of the coherence as functions of detuning $\delta=\omega_0-\omega$ and rescaled time $\Gamma t$ for both weak and strong couplings. One observes that when $\delta$ departures from zero, the decay of coherence speeds up achieving the fastest decay around $\delta=J$. It is interesting to highlight the role of the cavity-cavity coupling parameter $J$ as a benchmark for having the fastest decay during the dynamics under the non-resonant condition. For larger detuning tending to the dispersive regime ($\delta \gg \kappa$), the decay of coherence is instead strongly slowed down \cite{bellomo2010PhysScrManiscalco}. However, as shown in Fig.~\ref{fig:detuning2D}, stationary coherence is forbidden out of resonance when the cavity $C_2$ is perfect. Since our main aim is the long-time preservation of quantum coherence and thus of entanglement, in the following we only focus on the condition of resonance between qubit and cavity frequencies.

\subsection*{Two-qubit entanglement preservation}

\begin{figure}[tbp]
\centering
\includegraphics[width=0.46\textwidth]{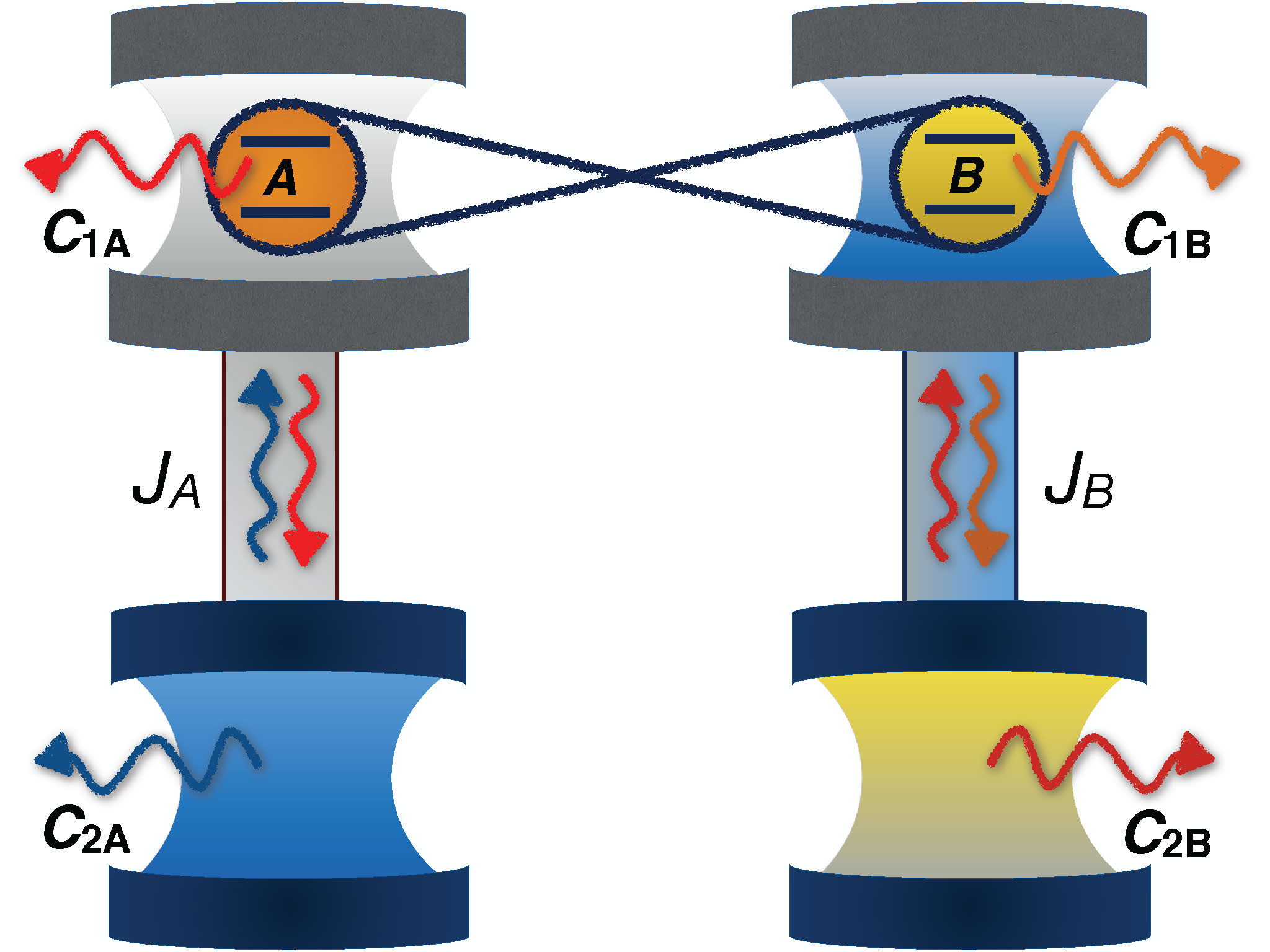}
\caption{\textbf{Scheme of the two-qubit architecture.} Two independent qubits $A$ and $B$, initially entangled, are locally embedded in a cavity $C_{1j}$ which is in turn coupled to a second cavity $C_{2j}$ by a coupling strength $J_j$ ($j=A,B$).}
\label{fig:system2}
\end{figure}

So far, we have studied the manipulation of coherence dynamics of a qubit via an
adjustment of coupling strength between two cavities. We now extend this architecture to
explore the possibility to harness and preserve the entanglement of two independent qubits, labeled as $A$ and $B$.
We thus consider $A$ ($B$) interacts locally with cavity $C_{1A}$ ($C_{1B}$) which is in turn coupled to cavity $C_{2A}$ ($C_{2B}$) with coupling strength $J_A$ ($J_B$), as illustrated in Fig.~\ref{fig:system2}. 
That is, we have two independent dynamics with each one consisting of a qubit $j$ ($j=A,B$) and two coupled cavities $C_{1j}$-$C_{2j}$.
The total Hamiltonian is then given by the sum of the two independent Hamiltonians, namely,
$H=\sum_{j}H_{j}$, where each $H_{j}$ is the single-qubit Hamiltonian of Eq. (\ref{singlequbithamiltonian}).
Denoting with $\Gamma _{1j}$ ($\Gamma _{2j}$) the decay rate of cavity $C_{1j}$ ($C_{2j}$), we shall assume $\Gamma_{1A}=\Gamma_{1B}=\Gamma$ as the unit of the other parameters.

As known for the case of independent subsystems, the complete dynamics of the two-qubit system can be obtained by
knowing that of each qubit interacting with its own environment \cite{bellomo2007PRL,bellomo2008PRA}.
By means of the single-qubit evolution, we can construct the evolved density matrix
of the two atoms, whose elements in the standard computational basis $\{\left|1\right\rangle\equiv\left|11\right\rangle,
\left|2\right\rangle\equiv\left|10\right\rangle,\left|3\right\rangle\equiv\left|01\right\rangle,
\left|4\right\rangle\equiv\left|00\right\rangle\}$ are
\begin{eqnarray}\label{two-at-evo}
\rho_{11}(t)&=&u_{t}^{A}u_{t}^{B}\rho_{11}(0) \nonumber \\ \nonumber
\rho_{22}(t)&=&u_{t}^{A}(1-u_{t}^{B})\rho_{11}(0)+u_{t}^{A}\rho_{22}(0)\\ \nonumber
\rho_{33}(t)&=&(1-u_{t}^{A})u_{t}^{B}\rho_{11}(0)+u_{t}^{B}\rho_{33}(0)\\ \nonumber
\rho_{44}(t)&=&(1-u_{t}^{A})(1-u_{t}^{B})\rho_{11}(0)+(1-u_{t}^{A})\rho_{22}(0)\\ \nonumber
&&+(1-u_{t}^{B})\rho_{33}(0)+\rho_{44}(0)\\ \nonumber
\rho_{14}(t)&=&\rho_{41}^{*}(t)=z_{t}^{A}z_{t}^{B}\rho_{14}(0)\\
\rho_{23}(t)&=&\rho_{32}^{*}(t)=z_{t}^{A}z_{t}^{B*}\rho_{23}(0),
\end{eqnarray}
where $\rho_{lm}(0)$ are the density matrix elements of the two-qubit initial state and $u_t^j$,$z_t^j$ are the time-dependent functions of Eq. (\ref{sing-at-evo}).

\begin{figure}[tbp]
\centering
{\includegraphics[width=0.47\textwidth]{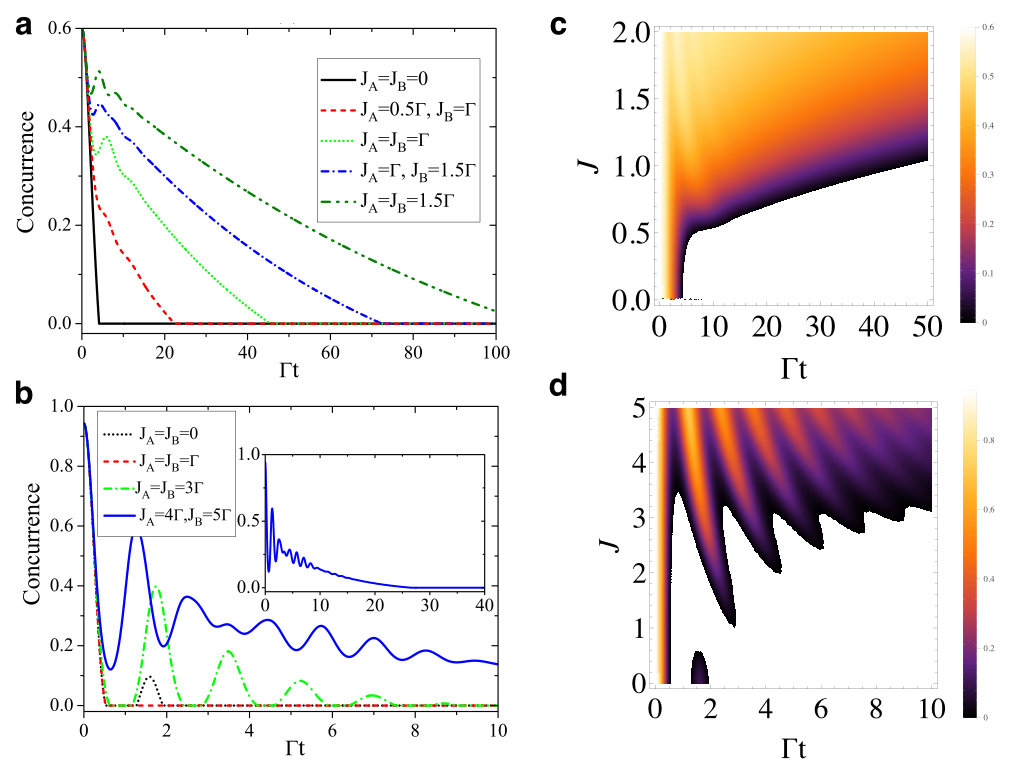}}
\caption{The dynamics of concurrence for different coupling strengths $J_{A}$ and $J_{B}$ in (a) weak qubit-cavity coupling regimes with $\kappa_{A}=\kappa_{B}=0.2\Gamma$ and (b) strong qubit-cavity coupling regimes with $\kappa_{A}=\kappa_{B}=2\Gamma$. The initial state weights are chosen as (a) $\alpha=\sqrt{1/10}$, $\beta=\sqrt{9/10}$ and (b) $\alpha=\sqrt{1/3}$, $\beta=\sqrt{2/3}$, while in both cases $\Gamma_{2A}=\Gamma_{2B}=0.2\Gamma$. The inset in (b) shows the long-time dynamics of concurrence for $J_{A}=4\Gamma$ and $J_{B}=5\Gamma$. Panels (c) and (d) show the density plots of the two-qubit concurrence as a function of $J$ ($J_{A}=J_{B}=J$ is here assumed) and scaled time $\Gamma t$, the others parameters being as in panels (a) and (b), respectively. The values of the concurrence in the density plots range within the interval: (c) $0$ \protect\includegraphics[width=2.2 cm]{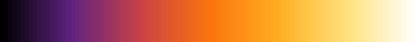} $0.6$; (d) $0$ \protect\includegraphics[width=2.2 cm]{legend2} $1$.}
\label{fig:con}
\end{figure}

We consider the qubits initially in an entangled state of the form $\left|\psi(0)\right\rangle=\alpha\left|00\right\rangle+
\beta\left|11\right\rangle$ ($|\alpha|^2+|\beta|^2=1$).
As is known, this type of entangled states with $|\beta|>|\alpha|$ suffers from
entanglement sudden death when each atom locally interacts with a dissipative environment \cite{santosPRA,yu2009Science,almeida2007Science}.
As far as non-Markovian environments are concerned, partial revivals of entanglement can occur \cite{lofrancoreview,bellomo2007PRL,bellomo2008PRA,lofranco2012PRA,LoFrancoNatCom,lofranco2012PhysScripta,darrigo2014IJQI,darrigo2013hidden,bellomo2008bell,ban1,ban2,liuPRA,yonac,manNJP,baiPRL,baiPRA} typically after asymptotically decaying to zero or after a finite dark period of complete disappearance. It would be useful in practical applications that the non-Markovian oscillations can occur when the entanglement still retain a relatively large value.
With our cavity-based architecture, on the one hand we show that the Markovian dynamics
of entanglement in the weak coupling regime between the atoms and the corresponding cavities (i.e., $C_{1A}$ and $C_{1B}$)
can be turned into non-Markovian one by increasing the coupling strengths between the cavities $C_{1A}$-$C_{2A}$
and (or) $C_{1B}$-$C_{2B}$; on the other hand, we find that the appearance of entanglement revivals can be shifted to earlier times.
We employ the concurrence \cite{Wootters98} to quantify the entanglement (see Methods), which for the two-qubit evolved state of Eq. (\ref{two-at-evo})
reads $\mathcal{C}_{AB}(t)=2\max\{0,|\rho_{14}(t)|-\sqrt{\rho_{22}(t)\rho_{33}(t)}\}$. Notice that the concurrence of the Bell-like initial state $\ket{\psi(0)}$ is $\mathcal{C}_{AB}(0)=2|\alpha\beta|$.
In Fig. \ref{fig:con}(a) we plot the dynamics of concurrence $\mathcal{C}_{AB}(t)$ in the weak coupling
regime between the two qubits with their corresponding cavities with $\kappa_{A}=\kappa_{B}=0.2\Gamma$
($\Gamma_{1A}=\Gamma_{1B}=\Gamma$ has been assumed). For two-qubit initial states
with $\alpha=\sqrt{1/10}$, $\beta=\sqrt{9/10}$, the entanglement experiences sudden death
without coupled cavities ($J_{A}=J_{B}=0$). By incorporating the additional cavities with relatively small coupling strength, e.g., $J_{A}=0.5\Gamma$ and $J_{B}=\Gamma$, the concurrence still undergoes a Markovian decay
but the time of entanglement disappearance is prolonged. Increasing the coupling strengths $J_{A}$, $J_{B}$ of
the relevant cavities drives the entanglement dynamics from Markovian regime to non-Markovian one. 
Moreover, the entanglement revivals after decay happen shortly after the evolution when the entanglement still has a large value.
In general, the concurrences are enhanced pronouncedly with $J_{A}$ and $J_{B}$. A comprehensive picture of the dynamics of concurrence as a function of coupling strength $J$ is shown in Fig. \ref{fig:con}(c) where we have assumed $J_{A}=J_{B}=J$.
In Fig. \ref{fig:con}(b) we plot the dynamics of $\mathcal{C}_{AB}(t)$ in the strong coupling
regime between qubit $j$ and its cavity $C_{1j}$ with $\kappa_{A}=\kappa_{B}=2\Gamma$ for which
the two-qubit dynamics is already non-Markovian in absence of cavity coupling, namely the entanglement can revive after dark periods. Remarkably, the figure shows that when the coupling $J_j$ between $C_{1j}$ and $C_{2j}$ is activated and gradually increased in each location, multiple transitions from non-Markovian to Markovian dynamics surface. We point out that the entanglement dynamics within the non-Markovian regime
exhibit different qualitative behaviors with respect to the first time when entanglement oscillates.
For instance, for $J_{A}=J_{B}=3\Gamma$, the non-Markovian entanglement oscillations (revivals) happen after its disappearance, while when $J_{A}=4\Gamma$ and $J_{B}=5\Gamma$ the entanglement oscillates before its sudden death. These dynamical features are clearly displayed in Fig. \ref{fig:con}(d).

\begin{figure*}[tbp]
\begin{center}
\includegraphics[width=0.8\textwidth]{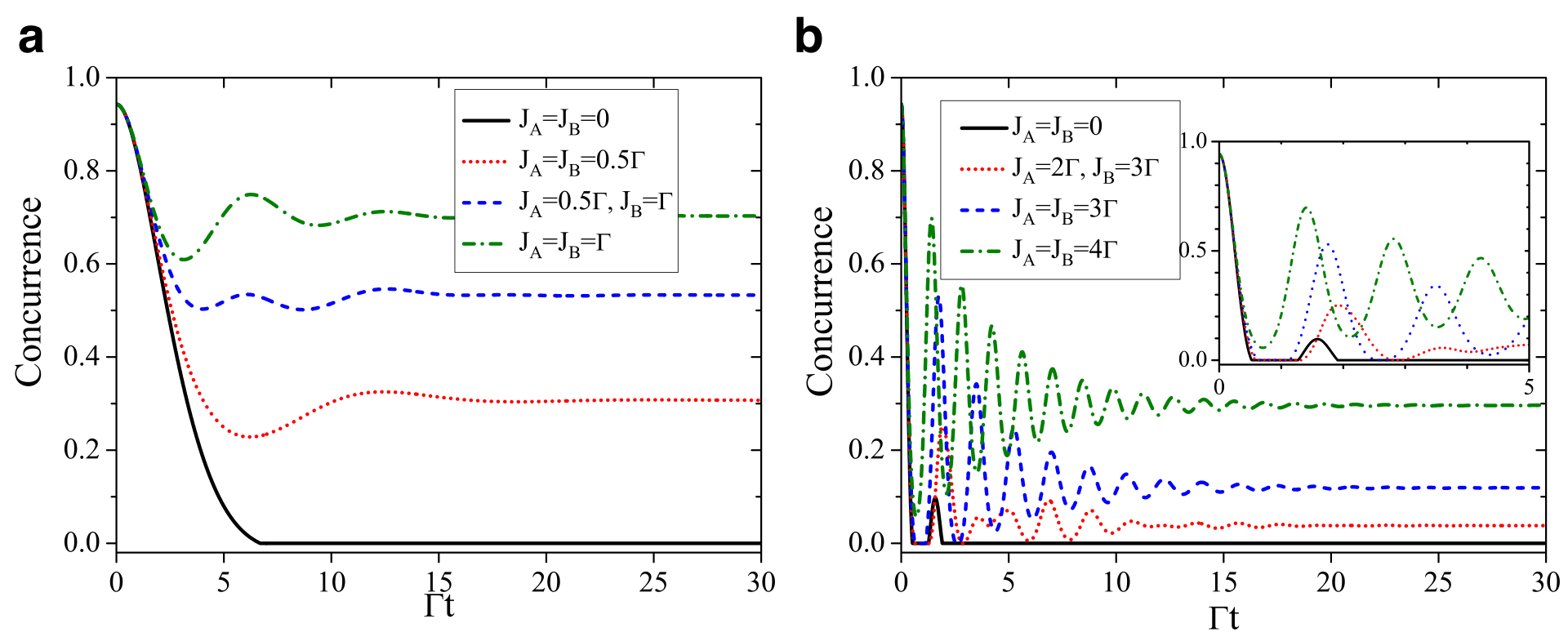}
\end{center}
\caption{The dynamics of concurrence for different coupling strengths $J_{A}$ and $J_{B}$
in the presence of ideal coupled cavities $C_{2A}$ and $C_{2B}$ with $\Gamma_{2A}=\Gamma_{2B}=0$ for
(a) $\kappa_{A}=0.2\Gamma$, $\kappa_{B}=0.3\Gamma$ and (b)
$\kappa_{A}=\kappa_{B}=2\Gamma$. The other parameters are chosen as $\alpha=\sqrt{1/3}$,
$\beta=\sqrt{2/3}$. The inset in (b) shows the short time dynamics of concurrence.}
\label{fig:con-trap}
\end{figure*}

\begin{figure}[tbp]
\begin{center}
\includegraphics[width=0.46\textwidth]{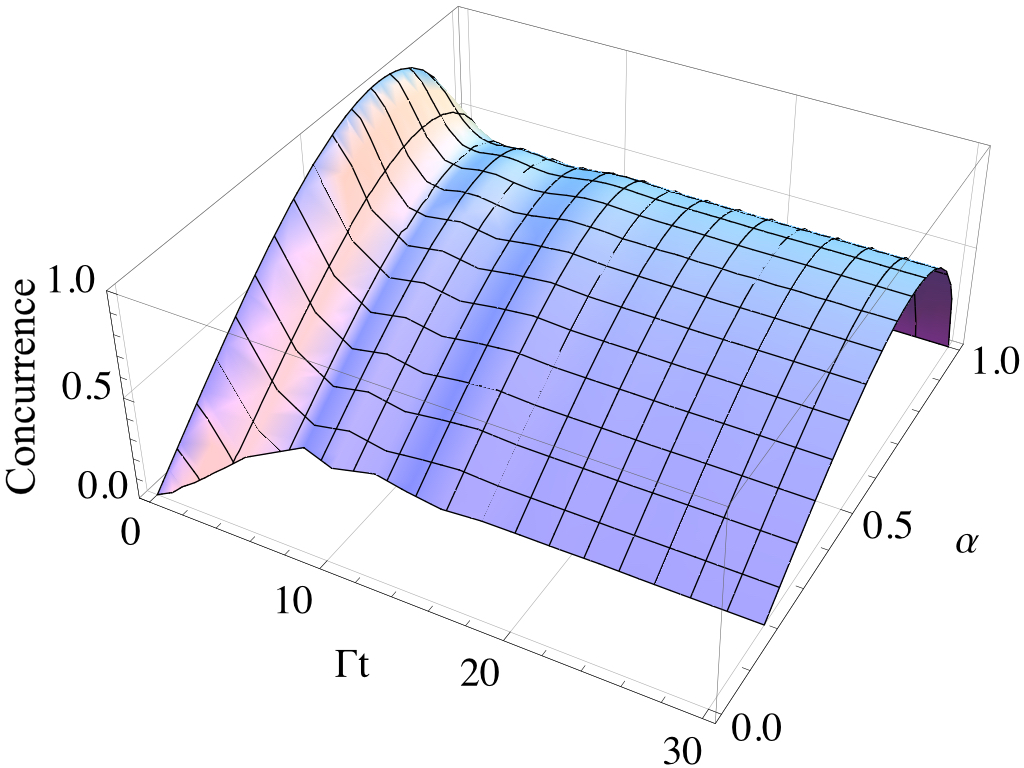}
\end{center}
\caption{The concurrence as a function of the two-qubit initial state parameter $\alpha$ and the scaled time $\Gamma t$
for $\kappa_{A}=0.2\Gamma$, $\kappa_{B}=0.3\Gamma$, $J_{A}=0.5\Gamma$, $J_{B}=\Gamma$ and $\Gamma_{2A}=\Gamma_{2B}=0$. The parameter $\alpha$ quantifies the initial entanglement according to the concurrence $\mathcal{C}_{AB}(0)=2|\alpha\beta|=2|\alpha|\sqrt{1-|\alpha|^2}$.}
\label{fig:initial}
\end{figure}

As expected according to the results obtained before on the single-qubit coherence, a steady concurrence arises in the long-time limit if the secondary cavities $C_{2A}$, $C_{2B}$ do not lose photons, i.e., $\Gamma_{2A}=\Gamma_{2B}=0$.
Fig. \ref{fig:con-trap}(a) shows the dynamics of concurrence for qubits coupled to their cavities with strengths $\kappa_{A}=0.2\Gamma$, $\kappa_{B}=0.3\Gamma$. We can readily see that, in absence of coupling with the secondary cavities ($J_{A}=J_{B}=0$),
the entanglement disappear at a finite time without any revival.
Contrarily, if the local couplings $C_{1j}$-$C_{2j}$ are switched on and increased, the entanglement does not vanish at a finite time any more and reaches a steady value after undergoing non-Markovian oscillations. Furthermore, the steady value of concurrence is proportional to the local cavity coupling strengths
$J_{A}$, $J_{B}$.
In Fig.~\ref{fig:con-trap}(b), the concurrence dynamics for $\kappa_{A}=\kappa_{B}=2\Gamma$ is plotted under which the two-qubit entanglement experiences non-Markovian features, that is revivals after dark periods, already in absence of coupled cavities, as shown by the black solid curve for $J_{A}=J_{B}=0$. Of course, in this case the entanglement eventually decays to zero. On the contrary, by adjusting suitable nonzero values of the local cavity couplings a considerable amount of entanglement can be trapped. As a peculiar qualitative dynamical feature, we highlight that the entanglement can revive and then be frozen after a finite dark period time of complete disappearance (e.g., see the inset of Fig.~\ref{fig:con-trap}(b), for the short-time dynamics with $J_{A}=2\Gamma$, $J_{B}=3\Gamma$ and also $J_{A}=J_{B}=3\Gamma$). 
We finally point out that the the amount of preserved entanglement depends on the choice of the initial state (i.e., on the initial amount of entanglement) of the two qubits. As displayed in Fig.~\ref{fig:initial}, the less initial entanglement, the less entanglement is in general maintained in the ideal case of $\Gamma_{2A}=\Gamma_{2B}=0$. However, since there is not a direct proportionality between the evolved concurrence $\mathcal{C}_{AB}(t)$ and its initial value $\mathcal{C}_{AB}(0)$, the maximal values of concurrence do not exactly appear at $\alpha = 1/\sqrt{2}$ (corresponding to maximal initial entanglement) at any time in the evolution, as instead one could expect. It can be then observed that nonzero entanglement trapping is achieved for $\alpha>0.2$.

\subsubsection*{Experimental paramaters}
We conclude our study by discussing the experimental feasibility of the cavity-based architecture here proposed for the two-qubit assembly. Due to its cavity quantum electrodynamics characteristics, our engineered environment finds its natural realization in the well-established framework of circuit quantum electrodynamics (cQED) with transmon qubits and coplanar waveguide cavities \cite{blaisPRA,steffenarxiv,schoelkopfarxiv,leekPRL,finkNature}. The entangled qubits can be initialized by using the standard technique of a transmission-line resonator as a quantum bus \cite{blaisPRA,dicarloNature}. Initial Bell-like states as the one we have considered here can be currently prepared with very high fidelity \cite{dicarloNature}.
Considering up-to-date experimental parameters \cite{steffenarxiv,schoelkopfarxiv,leekPRL,finkNature} applied to our global system of Fig.~\ref{fig:system2}, the average photon decay rate for the cavity $C_{1j}$ ($j=A,B$) containing the qubit is $\Gamma_{1j} \in [1\ \mathrm{MHz},10\ \mathrm{MHz}]$, while the average photon lifetime for the high quality factor cavity $C_{2j}$ is $\tau_2\approx 55$ $\mu$s \cite{schoelkopfarxiv}, which implies
$\Gamma_{2j} \approx 10^{-2} \mathrm{MHz}\in [10^{-2} \Gamma_{1j},10^{-3} \Gamma_{1j}]$. The qubit-cavity interaction intensity $\kappa_j$ and the cavity-cavity coupling strength $J_j$ are usually of the same order of magnitude, with typical values $\kappa_j \sim J_j \in [1\ \mathrm{MHz},100\ \mathrm{MHz}]=[0.1 \Gamma_{1j},10 \Gamma_{1j}]$. The typical cavity frequency is $\omega \sim 2\pi \times 10$ GHz \cite{blaisPRA} while the qubit transition frequency can be arbitrarily adjusted in order to be resonant with the cavity frequency. 
The above experimental parameters put our system under the condition $\kappa_j \ll \omega $ which guarantees the validity of the rotating wave approximation (RWA) for the qubit-cavity interaction here considered in the Hamiltonian of equation~(\ref{singlequbithamiltonian}). 

In order to assess the extent of entanglement preservation expected under these experimental conditions, we can analyze the concurrence evolution under the same parameters of Fig.~\ref{fig:con-trap}(a) for $\kappa_j$, $J_j$, which are already within the experimental values, but with $\Gamma_{2A}=\Gamma_{2B}=\Gamma_2= 10^{-2} \Gamma, 10^{-3} \Gamma$ instead of being zero (ideal case), where $\Gamma=\Gamma_{1A} =\Gamma_{1B} \in [1\ \mathrm{MHz},10\ \mathrm{MHz}]$.
The natural estimated disappearance time of entanglement in absence of coupling between the cavities ($J_{A}=J_{B}=0$) is $\bar{t}=6.69/\Gamma \in [669 \ \mathrm{ns}, 6.69\ \mu\mathrm{s}]$, as seen from Fig.~\ref{fig:con-trap}(a).
When considering the experimental achievable decay rates for the cavities $C_{2j}$, we find that the entanglement is expected to be preserved until times $t^\ast$ orders of magnitude longer than $\bar{t}$, as shown in Table \ref{tab:times}.
In the case of higher quality factors for the cavities $C_{2j}$, such that the photon decay rate is of the order of $\Gamma_{2} = 10^{-4}\Gamma$, the entanglement can last even until the order of the seconds. These results provide a clear evidence of the practical powerful of our simple two-qubit architecture in significantly extending quantum entanglement lifetime for the implementation of given entanglement-based quantum tasks and algorithms \cite{obrienreview,dicarloNature,horodecki2009RMP,brunnerRMP}.

\begin{table*}[ht]
\centering
\begin{tabular}{|l || c | c |}
\hline
$\Gamma_2/\Gamma$ & $J_{A}/\Gamma=J_{B}/\Gamma=0.5$ & $J_{A}/\Gamma=0.5,\ J_{B}/\Gamma=1$
\\
\hline
$10^{-2}$ & $t^\ast=454/\Gamma \in [45.4 \ \mu\mathrm{s}, 454\ \mu\mathrm{s}]$ &
$t^\ast=974/\Gamma\in [97.4 \ \mu\mathrm{s}, 974\ \mu\mathrm{s}]$ \\
\hline
$10^{-3}$ & $t^\ast=4481/\Gamma\in [448 \ \mu\mathrm{s}, 4.48\ \mathrm{ms}]$ &
$t^\ast = 9686/\Gamma \in [0.967 \ \mathrm{ms}, 9.67\ \mathrm{ms}]$ \\
\hline
\end{tabular}
\caption{\label{tab:times} Estimates of the experimental entanglement lifetimes $t^\ast$ for different values of the second cavities decay rates $\Gamma_2$ and the local cavity couplings $J_A$, $J_B$. These values are to be compared with the natural entanglement lifetime without cavity coupling, $\bar{t}\in [669 \ \mathrm{ns}, 6.69\ \mu\mathrm{s}]$. The reference unit $\Gamma \in [1\ \mathrm{MHz},10\ \mathrm{MHz}]$.}
\end{table*}

It is worth to mention that nowadays cQED technologies are also able to create a qubit-cavity coupling strength comparable to the cavity frequency, thus entering the so-called ultra-strong coupling regime \cite{niemckzyk}. In that case the RWA is to be relaxed and the counter-rotating terms in the qubit-cavity interaction have to be taken into account. According to known results for the single qubit evolution beyond the RWA \cite{werlangNoRWA}, it appears that the main effect of the counter-rotating terms in the Rabi Hamiltonian is the photon creation from vacuum under dephasing noise, which in turns induces a bit-flip error in the qubit evolution. This photon creation would be instead suppressed in the presence of dissipative (damping) mechanisms \cite{werlangNoRWA}. Since our cavity-based architecture is subject to amplitude damping noise, the qualitative long-time dynamics of quantum coherence and thus of entanglement are expected not to be significantly modified with respect to the case when RWA is retained. These argumentations stimulate a detailed study of the performance of our proposed architecture under the ultra-strong coupling regime out of RWA, to be addressed elsewhere.

\section*{Discussion}

In this work, we have analyzed the possibility to manipulate and maintain quantum coherence and entanglement of quantum systems by means of a simple yet effective cavity-based engineered environment. In particular, we have seen how an environmental architecture made of two coupled lossy cavities enables a switch between Markovian and non-Markovian regimes for the dynamics of a qubit (artificial atom) embedded in one of the cavity. This feature possesses an intrinsic interest in the context of controlling memory effects of open quantum systems. Moreover, if the cavity without qubit has a small photon leakage with respect to the other one, qubit coherence can be efficiently maintained.

We mention that our cavity-based architecture for the single qubit can be viewed as the physical realization of a photonic band gap for the qubit \cite{laurapseudo}, inhibiting its spontaneous emission. This property, then extended to the case of two independent qubits locally subject to such an engineered environment, has allowed us to show that quantum entanglement can be robustly shielded from decay, reaching a steady-state entanglement in the limit of perfect cavities.
The emergence of this steady-state entanglement within our proposed architecture confirms the mechanism of entanglement preservation when the qubit-environment interaction is dissipative: namely, the simultaneous existence of a bound state between the qubit and its local environment and of a non-Markovian dynamics for the qubit \cite{non-Mar3}. We remark that this condition is here shown to be efficiently approximated within current experimental parameters such as to maintain a substantial fraction of the entanglement initially shared between the qubits during the evolution. Moreover, we highlight that this goal is achieved even if the local reservoir (cavity) embedding the qubit is memoryless, thanks to the exploitation of an additional good-quality cavity suitably coupled to the first one. Specifically, we have found that, by suitably adjusting the control parameter constituted by this local cavity coupling, the entanglement between the separated qubits can be exploited for times orders of magnitude longer than the natural time of its disappearance in absence of the cavity coupling. These times are expected to be long enough to perform various quantum tasks \cite{obrienreview,dicarloNature}.

Our long-living quantum entanglement scheme, besides its simplicity, is straightforwardly extendable to many qubits, thus fulfilling the scalability requirement for complex quantum information and computation protocols. The fact that the qubits are independent and noninteracting also allows for the desirable individual operations on each constituent of a quantum hardware.
The results of this work provide new insights regarding the control of the fundamental non-Markovian character of open quantum system dynamics and pave the way to further experimental developments towards the realization of devices able to preserve quantum resources.

\section*{Methods}

\subsection*{Functions of the single qubit density matrix}
Let us denote with $\mathcal{L}^{-1}\{ L(s) \}(t)$ the inverse Laplace transform of $L(s)$.
Then, the functions $u_{t}$ and $z_{t}$ appearing in Eq. (\ref{sing-at-evo}) are expressed as
\begin{equation}
 u_{t}=|z_{t}|^{2},\ z_{t} = \mathcal{L}^{-1}\{ F(s)/G(s)\}(t),\nonumber
 \end{equation}
where
\begin{eqnarray}
  F(s)&=& -4J^{2}-(2s+2i\omega+\Gamma_{1}) (2s+2i\omega+\Gamma_{2}),\\
   \nonumber
  G(s)&=&2 \kappa^2 (2 s + 2 i\omega + \Gamma_{2}) + [s + i (\delta + \omega)] \nonumber\\
  &&\times\{4 [J^2 + (s + i\omega)^2] +
    2 (s + i\omega) \Gamma_{2}  \nonumber\\
  &&+  \Gamma_{1} (2 s + 2 i\omega + \Gamma_{2})\}. \nonumber
  \end{eqnarray}

\subsection*{Entanglement quantification by concurrence}
Entanglement for an arbitrary state $\rho_{AB}$ of two qubits is quantified by concurrence \cite{amico2008RMP,Wootters98}
\begin{equation}
\mathcal{C}_{AB}=\mathcal{C}(\rho_{AB})=\textrm{max}\{0,\sqrt{\chi_{1}}-\sqrt{\chi_{2}}-\sqrt{\chi_{3}}-\sqrt{\chi_{4}}\},
\end{equation}
where $\chi_{i}$ ($i=1,\ldots,4$) are the eigenvalues in decreasing order of the matrix $\rho_{AB}(\sigma_{y}\otimes\sigma_{y})\rho_{AB}^{\ast}(\sigma_{y}\otimes\sigma_{y})$, with $\sigma_{y}$ denoting the second Pauli matrix and $\rho_{AB}^{\ast}$ corresponding to the complex conjugate of the two-qubit density matrix $\rho_{AB}$ in the canonical computational basis $\{\left|11\right\rangle,\left|10\right\rangle,\left|01\right\rangle,\left|00\right\rangle\}$.

\section*{Acknowledgements}
In this work Z.X.M. and Y.J.X. are supported by the National Natural Science Foundation (China) under Grants Nos. 11204156, 61178012 and 11247240,
and the Promotive Research Fund for Excellent Young and Middle-Aged Scientists of Shandong Province (China) under Project No. BS2013DX034.
R.L.F. acknowledges support by the Brazilian funding agency CAPES [Pesquisador Visitante Especial Grant No. 108/2012].

\end{document}